# ROUTING IN ALL-OPTICAL NETWORKS USING RECURSIVE STATE SPACE TECHNIQUE

Mohan Kumar S[1] and Jagadeesha S N[2]

[1]Ph.D, Research Center, Jawaharalal Nehru National College of Engineering,
shimoga, Karnataka, India.
[2]Department of Computer Science and Engineering,
Jawaharalal Nehru National College of Engineering, shimoga.

*ABSTRACT*

*In this papr, we have minimized the effects of failures on network performace, by using suitable Routing and Wavelenghth Assignment(RWA) method without disturbing other performance criteria such as blocking probability(BP) and network management(NM). The computation complexity is reduced by using Kalaman Filter(KF) techniques. The minimum reconfiguration probability routing (MRPR) algorithm must be able to select most reliable routes and assign wavelengths to connections in a manner that utilizes the light path(LP) established efficiently considering all possible requests.*

*KEYWORDS*

*Routing Wavelength Assignment (RWA), Blocking Probability(BP) Network Management(NM), Kalaman Filter(KF),Wavelength interchange(WI), Light Path(LP), Process Noise(PN), Least Loaded Routing(LLR), Minimum Reconfiguration Probability Routing (MRPR), Max-Sum Routing(MR), Adaptive Unconstrained Routing(AUR)*

## 1. INTRODUCTION

In this paper, we have considered a suitable algorithm for Routing and Wavelength Assignment(RWA) for wide area networks, such as Wavelength routing network (WRN) which has scalable architecture. WRN has mesh like structure consisting of links having one or more fibers at each input port to output port in the optical domain. The challenging problem in these networks are RWA and controlling problems. In these problems, provision of connections, called lightpaths in a scalable architecture usually span multiple links. Hence, light path may be assigned to different links along the route. This process is called Routing and Wavelength Assignment. In this process, lightpaths share one or more fibers links and different wavelengths. To establish a lightpath, a route should be discovered between source and destination and suitable wavelength need to be assigned to that route. Some of the commonly used performance criteria for Routing and Wavelength Assignment are throughput and blocking probability[10]. There are many algorithms proposed for this purpose with optimal solutions[10]. The main assumption of these algorithms are that, traffic volume is static for a long time period and these networks are reconfigured only to reflect changes in the long term traffic demand[1]. Although static demand has been reasonable assumption for voice data communication, in current trends and future, data intensive networks are rapidly changing. Therefore dynamic RWA algorithms which support request arrivals and lightpath terminations at stochastic/randaom times are needed. Hence predefined set of routes are searched in a predefined order to accommodate the request. Then the smallest index randomly selects a wavelength available on the route. If not, request is blocked. Usually one or more minimum hop routes are used and fixed order search is carried out without taking into account the congestion on the link. It will not further improve blocking performance. An adaptive RWA algorithm makes use of the network state information at the time of routing to find the optimum path using filter approach. Least loaded routing(LLR), Max-sum routing (MR) and Adaptive unconstrained routing(AUR) are the examples for adaptive RWA process[3].

An Optical switch has ability to minimize the effects of failures on network performance by using a suitable routing and wavelength assignment method without disturbing other performance criteria such as network management and blocking probability[4]. The main challenge in the Wavelength Routing Network (WRN) is the provision of connections called lightpaths between the users of network. This method is called Routing and Wavelength Assignment scheme (RWA)[2][10]. Here, one of the efficient Routing Wavelength Assignment method known as statistically predictive dynamic RWA algorithm[3] is implemented using the kalman filter. This algorithm makes use of the network state information at the time of routing to find the optimum path according to an objective function for the request. By choosing as much reliable router/links as possible in the RWA process, it is possible to minimize the mean number of light paths broken due to failure. However, considering only reliability characteristics, lightpaths may have to be routed on longer routes and blocking performance may be deteriorated. For this reason, this algorithm is based on the joint optimization of the probability of reconfiguration due to router/link failures and probability of blocking for the future requests. Therefore effect of potential router/link failures is minimized without disturbing blocking probability performance[7]. For this purpose lightpath arrival/holding time and failure arrival statistics collected for each link and router, as well as the network state information at the time of request arrival are used in routing decisions. That is, the behavior of the network is predicted by current state information and statistics of the past, to assign the most reliable path to the lightpath requests[9].

In the following section Minimum Reconfiguration Probability Routing (MRPR) algorithm is presented, then the cost function is derived to process RWA in WRN networks. Finally simulations studies are presented and analyzed in section 5.0. The concluding remarks and directions for future work are presented in section 6.0.

## 2. MINIMUM COST PATH FOR A LIGHTPATH

In Wavelength Interchange networks, the wavelength routers have wavelength converters at the output ports and are able to change the wavelength of all lightpath passing through it. Hence the blocking of requests due to wavelength conflicts can be avoided and the blocking probability can be significantly reduced[7]. Wavelength routers can also switch the wavelength of lightpaths, the RWA problem in WI networks reduce to the light path routing problem. The routing problem is solved, wavelengths on the links along the route can be assigned randomly. In order to find the route with minimum reconfiguration probability for a lightpath request, a simple auxiliary graph G= *(N,E)*,where nodes *N* represent the routers and each directed edge*(i,j)*, *E* represents the link *(i,j)* from router *i* to router *j*, is constructed. Then, cost of each edge*(i,j)* is set to:

$$C_{ij} = \begin{cases} -\ln(1-F_{ij}) - \ln(1-R_{ij}) - \ln(1-F_j), & \text{edge can not be used} \\ \infty, & \text{other wise} \end{cases} \quad\ldots\ldots\ldots\ldots(1)$$

Where, *Fij* is the probability of reconfiguration due to failure and *Rij* is the probability of reconfiguration due to repacking on the link from router *i* to router *j* and *Fj* is the probability of reconfiguration due to failure on router *j* for the lightpath to be routed [3]. In equation(1), Edge cannot be used' means that the link *(i,j)* has no free wavelength channel at the time of routing.

## 2.1 Reconfiguration Due to Failure

In this paper, we present an efficient routing algorithm based on Kalman Filtering techniques to predict a future state of the router in order to determine the accurate state of router to reconfigure due to link or route failure. we propose the usage of Kalman Filters in routing. First, Kalman Filter is employed for estimating the system based on several unknown parameters such as processes and measurement noises at routing device.The probability of reconfiguration due to failure for a lightpath on a resource (link or router) can be predicted in terms of mean of inter-arrival times on that resource and mean of holding time for the lightpaths between the source and destination routers. Probability of configuration due to failure, *F*, for a lightpath on a resource is equal to the probability of a failure on that resource during the lifetime of the lightpath[3]. *F* can be found as:

$$F = \int_{y=0}^{\infty}(\int_{x=0}^{y} f(x)dx)h(y)\,dy \dots\dots\dots\dots\dots\dots\dots\dots(2)$$

Where *x* is random variable representing failure inter-arrival times on the resource, *f(x)* is the probability distribution function (pdf) of random variable *x*, *y* is a random variable representing the lightpath holding times between the source and destination routers, *h(y)* is the pdf of random variable *y*.

To determine F, knowledge of f(x) and h(y) are required. A straightforward way to evaluate equation(2) is to approximate distribution function with mean and variance equal to the corresponding values obtained from the statistics. Lightpath holding times are usually approximated by an exponential distribution function and failure inter-arrival times are usually approximated by an exponential or a Weibull distribution function, which will closely approximate the observed phenomena. This calculation is done using Kalman equations(11) in order to reduce the computational complexity. In particular, if both failures inter arrival-time and lightpath holding times are approximated by exponential distribution functions, F can be found as:

$$F = \int_{y=0}^{\infty}(\int_{x=0}^{y} \frac{1}{m_f}e^{-\frac{x}{m_j\,dx}})\,\frac{1}{m_f}e^{-\frac{y}{m_h}ti\theta}\,dy = \frac{m_h}{m_h + m_f}\dots\dots\dots\dots\dots(3)$$

Where, $m_h$ and $m_f$ are the mean holding time and mean failure inter arrival time, respectively.

## 2.2 Reconfiguration Due to Repacking

The probability of reconfiguration due to repacking, R, for a lightpath on a link in the network can be predicted in terms of the number of lightpaths currently passing through the link, arrival rate and service time statistics for the lightpaths on the link[3]. To find R, we need to find the probability of a call blocked due to lack of free capacity on that link during the lifetime of the lightpath.

In order to evaluate the approximate repacking probabilities, we make following assumptions:

- Links in the network are independent of each other. That is, lightpath arrivals on each link are independent processes and a lightpath on an 'n' link route behaves like 'n' independent lightpaths.

- Poisson arrivals and exponential holding times: lightpaths arrive on alink under consideration according to a Poisson process with rate λ and lightpath holding times are exponentially distributed with mean 1/μ.

- The repacking probability on the link 'R' is independent of other links. Therefore, the repacking probability for a lightpath on route 'r' denoted as $R_r$ can be found from individual link repacking probabilities, $R_l$ as:

$$R_p = 1 - \prod_{l \in \rho}(1 - R_l) \quad\quad\quad\quad (4)$$

With the help of these assumptions, we can find the repacking probability 'R' for a lightpath request, $lp_0$ on a link with capacity C and having $N_0$ lightpaths ($N_0 < C$), can be by modeling as a Markov Process and is shown in figure1. In this process, all states except the one labeled by 'r' are transient states corresponding to the number of lightpaths on the link before any repacking is experienced. On the other hand, the state labeled by 'r' is the trapping state, which represents the repacking occurrence before the termination of lightpath.

$$\acute{p}_1(t) = \lambda p_1(t) + \mu p_2(t) \quad\quad (5)$$

$$\acute{p}_1(t) = -(\lambda + (n-1)\mu)p_n(t) + n\mu p_{n+1}(t) + \lambda p_{n-1}(t), \quad 1 < n < c, \quad (6)$$

$$\acute{p}_c(t) = -(\lambda + (c-1)\mu)p_c(t) + \lambda p_{c-1}(t) \quad\quad (7)$$

$$\acute{p}_c(t) = \lambda p_c(t), \quad\quad (8)$$

$$p_n(t) = \begin{cases} 1, & n = n_0 + 1 \\ x, & otherwise \end{cases} \quad 1 \le n \le c \quad p_r(0) = 0 \quad\quad (9)$$

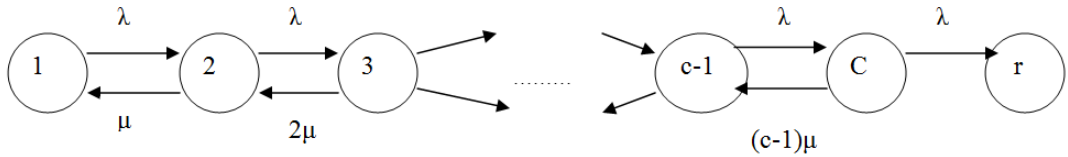

Figure 1: Constructed State diagram of Markov model to determine R

Where $p_n(t)$ for n= 1,2..C are the probability of having n light paths on the link at time 't' and no repacking has occurred up to time 't'. The $P_r(t)$ is the probability of experiencing a repacking on the link up to time t, and ($N_0$ +1) is the state of the link at time t=0, if lightpath is routed on this link. In this process, we assume that $lp_0$ remains in the link for t=>0, or equivalently, at least one lightpath exists in the link for t=>0. Therefore, the number of lightpaths that may terminate at state N equals N-1, and death rate at state N equals (N-1) μ. Finally the probability of experiencing repacking on the link until $lp_0$ terminates can be found by finding the expected value of $P_r(t)$ as:

$$R_l = \int_0^\infty P_r(t)\mu e^{-\mu t}dt \quad\quad (10)$$

Therefore, if we assume that when repacking occurs the lightpath to be re-routed is randomly selected lightpath in the link, we can find the probability of reconfiguration due to repacking for a lightpath as:

$$R = \frac{1}{C}\int_0^\infty P_r(t)\mu e^{-\mu t}dt \quad\quad (11)$$

After solving the equations(10) and (11), probability of the repacking can be calculated as:

$$R = \frac{E(C,\rho)}{C*E(n,\rho)} \dots\dots\dots\dots\dots\dots\dots\dots\dots\dots\dots\dots\dots\dots\dots\dots(12)$$

Where $\rho = \lambda/\mu$ and E (n, ρ) is the Erlang Loss Formula defined as:

$$E(n,\rho) = \frac{\rho^n/n!}{\sum_{i=0}^{n} \rho^i/i!} \dots\dots\dots\dots\dots\dots\dots\dots\dots\dots\dots\dots\dots\dots(13)$$

As a result, we find the path reconfiguration probabilities in terms of router or link failure and link repacking probabilities as:

$$C_{ij} = 1 - \prod_{(i,j)\in\rho} (1-F_{ij})(1-R_{ij})(1-F_j) \dots\dots\dots\dots\dots\dots\dots\dots\dots(14)$$

Where $C_{ij}$ is the cost of lightpath between nodes i and j. $F_{ij}$ is the probability of reconfiguration due to failure in the link between node i and j. $R_{ij}$ is the probability of repacking between the nodes i and j and $F_j$ is the pro bability of reconfiguration due to the failure in the router located at node j [8][9].

## 3. KALMAN FILTER TECHNIQUE

Kalman filter can be employed as a alternative to the Markov model to estimate the arrival and holding times in an WRN and WI networks and solve the RWA problem are to be estimated.

### 3.1 Estimating the Arrival Time at Nodes of Stochastic and Dynamic Networks Using Kalman Filter Technique

To estimate the arrival times at the nodes of a stochastic and dynamic network[5][7] a step prior to route planning. A algorithm is developed to predict the traveling times along the arcs and estimate the arrival times at the nodes of the network in real-time. It is shown that, under fairly mild conditions, the developed arrival time estimator is unbiased and that the error variance of the estimator is bounded.

Given a directed graph G = (N, A), with |N| = n and |A| = m, in dynamic problems, a non-negative travel time $d_{ij}(t)$ is associated with each arc(i, j) with the following meaning:

if *t* is a feasible leaving time from node i along the arc(i, j), then $t + d_{ij}(t)$ is the arrival time at node j. In addition to the travel time, a time-dependent travel cost $c_{ij}(t)$ can be associated with (i, j), which is the cost of traveling from i to j through (i, j) starting at time t. There is the possibility of waiting at the nodes; in this case, a (unit time) waiting cost $w_i(t)$ can be associated with node i, which gives the cost of waiting for unit time at i at time t. Given a route in a dynamic stochastic transportation network, we develop a methodology to estimate the arrival times at the nodes of that route. To estimate the arrival times, first step is developing a technique to predict the traveling times on the arcs of the network in real-time. In this technique, available historical data are used for predicting the traveling times, and new measurements are used to correct and update our prediction at each instant of time. More precisely, this proposed methodology consists of the following two stages:

    1. Predicting traveling times on arcs: Given the time of the day together with the historical and real-time data of traveling times on arcs of a transportation network, we predict the future traveling times on those arcs, recursively.

2. Estimating arrival times at nodes: Given a route in the network, the departure time from the first node of the route, and the predicted traveling times on arcs of the network (found in stage 1), the arrival times at the nodes of the route are estimated.

In the following section, each stage is investigated and discussed in detail.

### 3.2 Predicting the Traveling Times on Arcs :

Let G:(N,A) be a transportation network (graph) with node set N={1...i, 1...j} and arc set A=(i,j). A typical transportation network is shown in Figure(2).

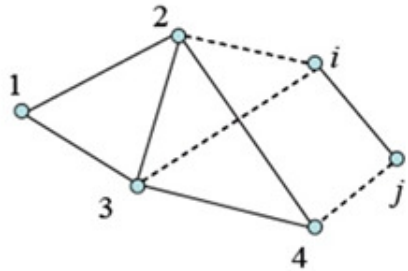

Figure 2: A typical transportation network

In Figure(2) Solid lines represent the direct connections (arcs) between two adjacent nodes and dashed lines are indirect connections, which consist of two or more arcs.

It is assumed that the historical data as well as the real-time information of traveling times on arcs of the network are available. Let  be the length of the planning horizon, K be an index of time in the planning horizon, $X_{ij}(k)$ be the traveling time between nodes i and j at time k, $X^h_{ij}(k)$ be the historical traveling time on arc (i,j) at time k, $U_{ij}(k)$ be the historical change in the traveling time on arc (i, j) from time k to k+1, $Y_{ij}(k)$ be the measured traveling time on arc (i, j) at time k.

The dynamic behavior of the traveling time on arc (i ,j) is given by the system model equation as:

$$X_{ij}(k + 1) = X_{ij}(k) + U_{ij}(k) + W_{ij}(k) \ldots\ldots\ldots\ldots\ldots\ldots\ldots\ldots\ldots (15)$$

$$Y_{ij}(k) = X_{ij}(k) + V_{ij}(k) \ldots\ldots\ldots\ldots\ldots\ldots\ldots\ldots\ldots\ldots (16)$$

Where $W_{ij}(k)$ is the traveling time disturbance on arc (i, j) at time k which is caused by the addition of white Gaussian noise during switching from one wavelength to another wavelength at each node. This noise is Known as process noise and $V_{ij}(k)$ is the error in the traveling time measurement of arc (i, j) at time k which is caused by the delay in the measurement. This delay in measurement is known as measurement noise. $W_{ij}(k)$ represents the real-time changes in the traveling time at time k, which are not included in the historical data $U_{ij}(k)$[6].

Here $U_{ij}(k)$, $W_{ij}(k)$, $V_{ij}(k)$ and $X_{ij}(0)$ are all mutually uncorrelated Gaussian random variables with the following specifications:

$$E\{U_{ij}(k)\} = \tau_{(ij)}(k), \quad E\{U_{ij}(k), U_{ij}(l)\} = \begin{cases} \sigma_{ij}^2(k) += \tau_{ij}^2(k) & k = 1 \\ 0, & \\ k! = 1 \end{cases} \ldots\ldots(17)$$

$$E\{W_{ij}(k)\} = 0;; , \quad E\{W_{ij}(k), W_{ij}(l)\} = \begin{cases} \sigma_{ij}^2(k) & k = 1 \\ 0 & k! = 1 \end{cases} \quad \text{...............(18)}$$

$$E\{W_{ij}(k)\} = 0;; , \quad E\{V_{ij}(k), V_{ij}(l)\} = \begin{cases} \sigma_{ij}^2(k) & k = 1 \\ 0 & k! = 1 \end{cases} \quad \text{................(19)}$$

### 3.3 Single Stage Predictor

In the single stage predictor case, given the measured traveling time $Y_{ij}(k)$ on arc (i, j) at time k, traveling time estimation $\hat{x}_{ij}(k + \frac{1}{k})$ is done on the arc (i, j) at the time k+1. The estimator that minimizes the mean squared error of the estimation is given as:

$$n\, \hat{x}_{ij}\left(k + \frac{1}{k}\right) = E\{X_{ij}(k+1)|\, Y_{ij}(k)\} \quad \text{.......................(20)}$$

From the equation (20), $\hat{x}_{ij}(k + \frac{1}{k})$ denotes the estimate of the travel time $X_{ij}(k+1)$ given measurement $Y_{ij}(k)$ and is called the mean-squared predicted estimator of $X_{ij}(k+1)$ Using designed dynamical model in equation 15, the predicted estimator $\hat{x}_{ij}(k + \frac{1}{k})$ can be written as:

$$\hat{x}_{ij}\left(k + \frac{1}{k}\right) = E\{X_{ij}(k) + U_{ij}(k) + W_{ij}(k)|Y_{ij}(k)\} \quad \text{.......................(21)}$$

$$= \hat{x}_{ij}(k|k) + \tau_{ij}(k) \quad \text{..................................................(22)}$$

Where $\hat{x}_{ij}(k|k)$ is the mean squared filtered estimator of $X_{ij}(k)$ given $Y_{ij}(k)$. In equation (22) indicates that to calculate the predicted estimator $\hat{x}_{ij}\left(k + \frac{1}{k}\right)$ the value of the filtered estimator $\hat{x}_{ij}(k|k)$ should be obtained. The mean and variance of the single stage predictor error of the mean-squared predicted estimator of $x_{ij}(k+1)$ is calculated. Finally single stage predictor is extended to $m^{th}$ stage predictor, with the measured traveling time $Y_{ij}(k)$ at time k on arc (i, j), to determine an unbiased estimate of traveling time on that arc at time k +m, where m>=1. This design part gives the traveling time estimation for the single stage between two nodes in the network.

### 3.4 The M$^{th}$ Stage Predictor

Similar to the single-stage predictor, estimator that minimizes the mean-squared estimation is given as:

$$\hat{x}_{ij}\left(k + \frac{m}{k}\right) = E\{X_{ij}(k+m)|Y_{ij}(k)\} \quad \text{...........................(23)}$$

Where $\hat{x}_{ij}\left(k + \frac{m}{k}\right)$ is the estimate of $X_{ij}(k+m)$ given the measurement $Y_{ij}(k)$. The mean and error covariance of m$^{th}$ stage predictor is calculated. The predicted estimate of traveling time $X_{ij}(k+m)$ on arc (i, j) at time k+m, m>=1, depends on the value of the filtered estimator $\hat{x}_{ij}(k/k)$ which is estimate of traveling time $\hat{x}_{ij}(k)$ at time k, given the measured traveling time $Y_{ij}(k)$ at time k, The predictor-corrector form of Kalman filter is used here to calculate the filtered estimator $\hat{x}_{ij}(k/k)$ as follows:

$$\hat{x}_{ij}(k/k) = \hat{x}_{ij}(k/k - 1) + K_{ij}(k)|Y_{ij}(k) - \hat{x}_{ij}(k/k - 1) \quad \text{..............(24)}$$

Where $K_{ij}(k)$ is the Kalman gain of arc(i,j) at time k is calculated to update the previous measurements. Therefore, at each time k and by using the mth stage predictor, the traveling time $\hat{x}_{ij}\left(k + \frac{m}{k}\right)$ on each arc (i,j) at time k+m can be predicted.

In the following section, Kalman filtering corrector-predictor technique is used to estimate the future traveling times on the arcs of graph G as shown in Figure(3).

### 3.5 Estimating Arrival Times at Nodes:

In the previous section, a technique based on Kalman filter to predict the traveling times on arcs of a given transportation network is explored. In this section, predicted traveling times on arcs is required to estimate the arrival times at the nodes of the network. Consider the graph 'G' as shown in figure(3).

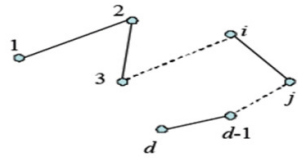

Figure 3 : A typical route *r* in the graph G

The route 'r' defined in graph 'G' as a sequence of nodes visited in the specified order. In Figure(3) shows a typical route 'r', which for convenience is represented as an ordered set i.e., R ={1, 2, i, j,.., d}. Let Ar be the arc set associated with route r.

Assumption is made such that the departure time from node 1 on route r, denoted by $Z^r{}_1$ , is given using the developed technique in previous Section, the predicted traveling times on arcs of graph 'G' are available. Given $Z^r{}_1$ , the arrival times at all the other nodes on route r in Figure(3) can be determined using the following set of equations.

$$Z_2^r = Z_1^r + X_{12}(Z_1^r) \quad \text{...............................................(25)}$$

$$Z_3^r = Z_2^r + X_{23}(Z_{12}^r) \quad \text{...........................................(26)}$$

$$.....$$
$$.....$$

$$Z_j^r = Z_i^r + X_{12}(Z_i^r) \quad \text{...........................................(27)}$$

$$....$$
$$....$$

$$Z_d^r = Z_{d-1}^r + X_{1d-1,d}(Z_{d-1}^r) \quad \text{...............................(28)}$$

Where $Z_i^r$ is the arrival time at node i taking route r, and $X_{12}(Z_i^r)$ is the traveling time on arc (i, j) at time $Z_i^r$ . Assumption is made such that the departure time from node i is equal to the arrival time at that node. This is achieved by assuming that there is no service time associated with the nodes of network G.

### 3.6 Estimating the Holding Time at Nodes of Dynamic Stochastic Networks

Given a directed graph G = (N, A), with |N| = n and |A| = m, in dynamic problems, a non-negative travel time $d_{ij}(t)$ is associated with each arc (i, j) with the following meaning: if t is a feasible leaving time from node j along the arc (i, j), then t + $d_{ij}(t)$ is the holding time at node j. In addition to the travel time, a time-dependent travel cost $C_{ij}$ (t) can be associated with (i, j),

which is the cost of traveling from i to j through (i, j). Mean holding time is calculated by estimating the traveling time between the source and destination nodes and prior knowledge of departure time at that node. Then $ij^{th}$ Kalman filter coefficients are calculated on the basis of above design. The implementation part is described in the following section.

## 4. PROPASED SYSTEM TO ESTIMATE AND MEASURE NOISE

Noises are random background events which have to be dealt with in every system processing real input signals (requests). They are not part of the ideal signal and may be caused by an effect of neighboring sources or delay in processing the request. The characteristics of noise depend on their source.

### 4.1 Implementation of Process Noise to Estimate The Unknown Parameters:

The process noise is generated because of the signals being processed on the same wavelength channel for some random period. Process noise can be analyzed both in wavelength convertible networks as well as networks without converters. Q (the model/input noise covariance) contributes to the overall uncertainty of the estimate as it is added to P (the error covariance matrix) in each time step. When Q is large the Kalman Filter large changes in the actual output more closely. This means there is a performance trade-off between tracking and noise in the output in the choice of Q for the Kalman Filter[6][11].

### 4.2 Implementation of Measurement Noise to Estimate the Unknown Parameters:

The measurement noise is added to the signal because of the delay in the measurement device. If the measuring device fails to measure the input data at some fixed time, then there will be delay. Data will be the lightpath request. If the lightpath request is blocked for more than its lifetime, then it is rejected permanently.

R (the measurement noise covariance) determines how much information from the sample is used. If R is high the Kalman Filter measurement isn't very accurate. When R is smaller the Kalman Filter output will follow the measurements more closely and accept more information from them.

The effect of P (the error covariance matrix) on the Kalman Filter estimate is that when P is small the Kalman Filter incorporates a lot less of the measurement into the estimate as it is fairly certain of its time. Ideally P gets as close to zero as possible to indicate that the model is accurate. P is generally reduced by measurements received; as there is more confidence in the estimated state if there is a measurement to confirm it. However the reduction of P is limited by the model/input variable error covariance Q which is added at each time step. Both P and R are incorporated into the Kalman Filter through the Kalman gain K. The value of K determines how much of the innovation (the difference between the actual measurement and the model measurement) is used to correct the estimate. K varies in proportion to the error covariance matrix P and is inversely proportional to the measurement covariance matrix R. If the measurement noise covariance R is large compared to the error covariance matrix P then the Kalman gain K will be small. This means the certainty of the measurement is small relative to the certainty of the current state model and the old model is better compared to the new measurement so that minimal adjustment to the estimate is required. Alternatively, if P is large compared to R, K will be large and the estimate of X is updated to look more like the measurement than the previous estimate. The innovation is weighted more heavily.

## 4.3 Important Routines in the Kalman Filter Implementation

In order to implement the Kalman Filter, the following steps need to be performed:

a) Model the System: A model of the system needs to be found. A knowledge of state space representations is useful here. Modeling the system essentially means working out the matrices Ô, B and H. Ô is the relationship between x in one time step, and x in the next, given no inputs. B is the relationship between the inputs and the state. H is the relationship between the measurement and the state (slightly more complicated for the Extended Kalman Filter).

b) Noise Parameters: In theory R and Q can be calculated directly from the real world where Q relates to model errors and input data errors, and R relates to measurement errors. In practice these can't always be obtained accurately (or the effort involved is too great), or the correct values don't give the required result. Tuning R too small will place too much emphasis on the measurements making the filter varying. Tuning Q too large gives the same result. Tuning R too large and Q too small will have the effect of making the filter too slow, and it won't keep up with the actual changes in x. These values can be adjusted until the filter gives the desired performance.

c) Initial Estimates: The state vector x and the error covariance matrix P need initial estimates. Any 'guess' for x will mean that the estimate will eventually converge on the right value, as long as P is non-zero. Given this, the best initial estimate is the 'middle' of where x is likely to be. P needs to be chosen significantly large so that the filter is not too slow and small enough that P doesn't remain large for too long.

d) Implement the Filter: Use the values obtained from the above steps and substitute them into the Kalman Filter equations.

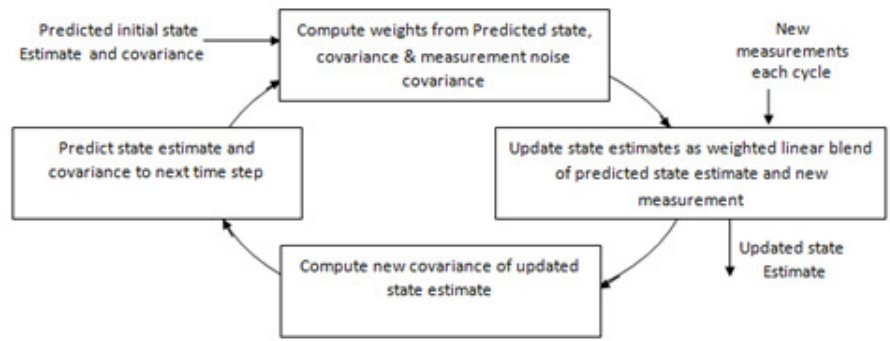

Figure 4: The Kalman Filter as a recursive linear filter.

At each cycle, the state estimate is updated by combining new measurements with the predicted state estimate from previous measurements. Figure(5) shows the Kalman filter algorithm and its four steps for computations are Gain computation, State estimate update, Covariance update, and Prediction[6].

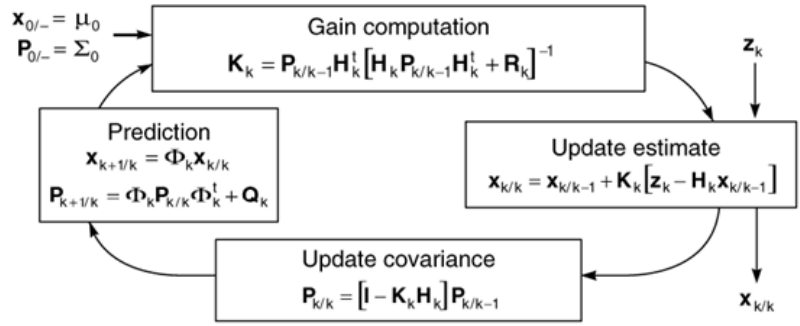

Figure 5: The Kalman filter algorithm updation and prediction.

### 4.4 Reconfiguration of Lightpath Using Erlang's Formula:

Probability of configuration due to failure, F, for a lightpath on a resource is equal to the probability of a failure on that resource during the lifetime of the lightpath. F can be found as:

$$F = \frac{m_h}{m_h + m_f} \quad \text{..................................................(29)}$$

From the above equation(29). Where $m_h$ and $m_f$ are the Mean holding time and mean failure inter arrival time, respectively. The Probability of the repacking can be calculated as follows:

$$R = \frac{E(c,\rho)}{c * E(n,\rho)} \quad \text{..................................................(30)}$$

Where $\rho = \lambda/\mu$ and E(n,ρ) is the Erlang's Loss formala defined as follows:

$$E(n, \rho) = \frac{\rho^n / n!}{\sum_{i=0}^{n} \rho^i / i!} \quad \text{..................................................(31)}$$

As a result, we find the path reconfiguration probabilities in terms of (router and link from eqution(14) failure and link repacking probabilities are as follows :

$$C_{ij} = 1 - \prod_{(i,j) \in \rho} (1 - F_{ij})(1 - R_{ij})(1 - F_j) \quad \text{.........................(32)}$$

## 5. SIMULATION RESULTS

we studied offline state space technique using the two well known approaches, namely, M/M/1 queuing and Kalman Filter state space model from the assumed topology. Our investigation reveals that the proposed state space algorithm is suitable for large networks.

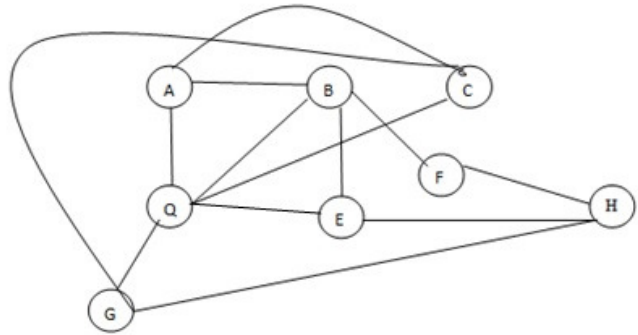

Figure 6: Assumed Topology

The assumed topology consiste of 13 Links and 6 rounters from A to H shown in the Figure (6). All the links are unidirection light paths and are established between the source 'A' to 'H' for each request. A Light path request arrive at the time for each request as a poisson process and lightpat holding times are considered to be exponential process. In simulation, each link in the assumed network is composed of 3 fibers with channels on each fiber link and routers have no converter. The MRPR algorithm used in the networks tries to route the lightpath on the route of all the possible routes. The blocking probability performance of MRPR are kept out of the scope. The abilities of Kalman filter based approach over the Markov model for estimation of mean holding time, average waiting time, inter-arrival time and their corrected time, are presented using computer simulation.

The Table(1) shows the recorded values for service begin, server idle time, service end, mean holding time and inter arrival time for M/M/1 queue size of 10. The number of customers' participating has been assumed to be 50.

Table 1: Shows Inter arrival time for the M/M/1 Markov model.

| Clock | Inter Arraiaval Time in µ sec | Next Arriaval time in µ sec | Service begin in µ sec | Service time in µ sec | Service end in µ sec | Service idle time in µ sec | Customer waiting time in µ sec |
|---|---|---|---|---|---|---|---|
| 0.06 | 0.02 | 0.09 | 0.06 | 2.04 | 2.10 | 0.06 | 0.00 |
| 0.09 | 0.20 | 0.29 | 0.06 | 0.06 | 2.10 | 0.06 | 0.00 |
| 0.29 | 2.64 | 2.93 | 0.06 | 2.04 | 2.10 | 0.06 | 0.20 |
| *2.10* | *2.64* | *2.93* | *2.10* | *1.22* | *3.33* | *0.00* | *3.63* |
| 2.93 | 4.62 | 7.55 | 2.10 | 1.22 | 3.33 | 0.00 | 2.64 |
| 3.33 | 4.62 | 5.55 | 3.33 | 1.09 | 4.42 | 0.00 | 0.80 |

From the Table(1) it is seen that, the total elapsed time is nothing but a interval between a service begin and service end for total request, i.e the total amount of Elapsed Time is equal to 9.67 µsec. The Average waiting time is equal to 1.216 µsec per arrival request. The Average Server idle time/request is equal to 0.18 µsec.

Table 2: Shows Inter arrival time for the Kalman filter model.

| No. Of iteration | Actual time in µ sec | Measured time in µ sec | Corrected time in µ sec | Estimated time in µ sec |
|---|---|---|---|---|
| 1 | 2.0 | 2.22 | 2.1935 | 2.2451 |
| 2 | 2.22 | 2.4135 | 2.3846 | 2.4379 |
| 3 | 2.413 | 2.6404 | 2.5734 | 2.6284 |
| *4* | *2.6404* | *2.7934* | *2.7600* | *2.8167* |
| 5 | 2.7934 | 2.9800 | 2.9444 | 3.0027 |

The effectiveness of using Kalman filter at each node to estimate the mean holding time, inter-arrival time and their corrected time, are shown in Table 2. As a result, it can be seen from the Tables(1) and (2), in the case M/M/1 from Table(1) for clock recorded of beginning time 2.10 µsec, It is seen that service begins at 2.10 µsec and Service ends at 3.33 µsec, and the customer waiting time is 3.63 µsec. However in the case of Kalman filter from the Table (2) the actual time is 2.6404 µsec. The measured time is 2.7934 µsec, and corrected time is 2.7600 µsec, which is for forth iteration. Hence waiting time is caluculated as measured time minus actual time 0.15774 µsec, per arrival request. For all optical networks, comparing, the delay experienced in state space technique using kalman filter approach is optimally better compared with markovian model.

Signal & Image Processing : An International Journal (SIPIJ) Vol.7, No.2, April 2016

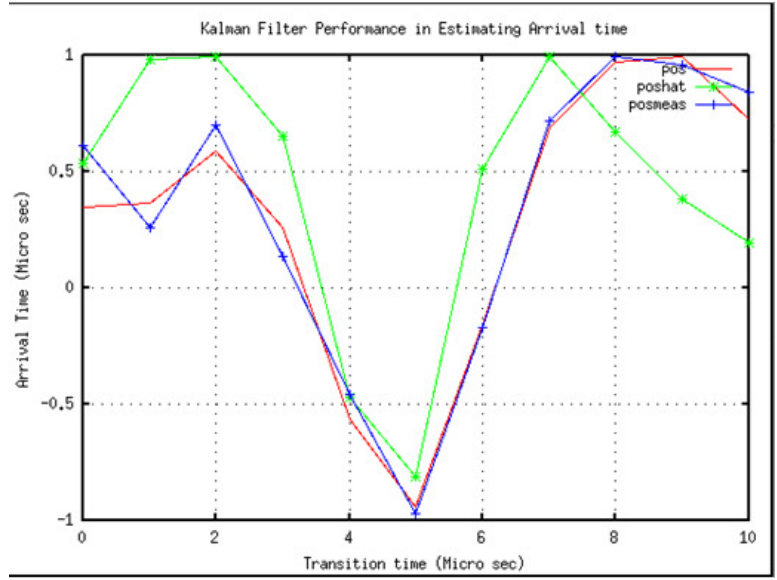

Figure 7: Kalman Filter analysis

The Transition time Vs. Arrival Time results has been plotted, in Figure(7), The red colour line shows the Actual value of the arrival time, the blue colour line indicates the measured value of Arrival time and green colour line shows the estimated values of Arrival time at different intervals of time when the measured noise is 0.02 and process noise is 0.01 using kalman filter technique. Since measured noise and process noise is small, the Kalman filter technique is able to retrieve the state of node.

Estimated arrival time ,when measurement noise=0.1 and process noise=0.1 has been considered to plot the graph of transition time Vs Arrival time.

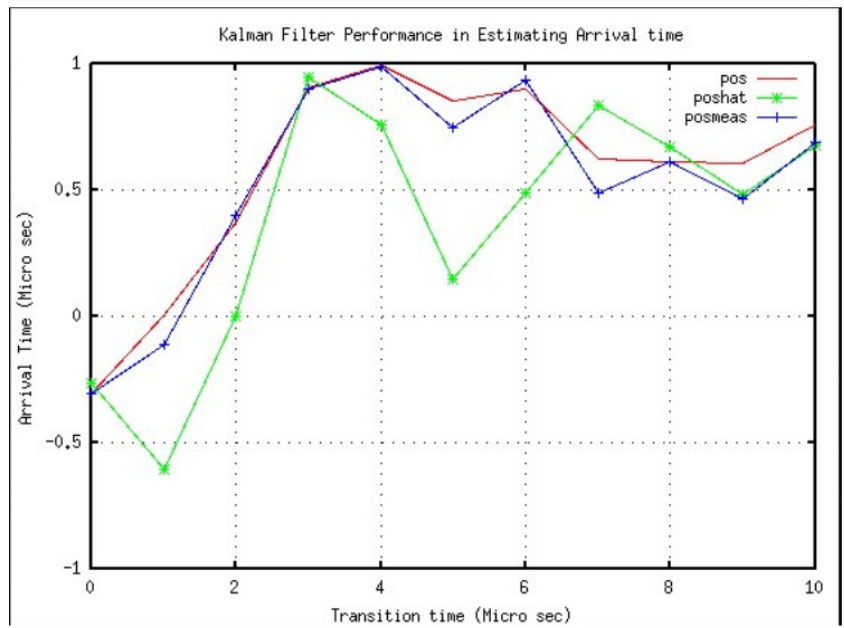

Figure 8: Performance in Estimating Arrival Time

As a second example, Figure(8) shows the performance in estimating the arrival time with a measurement noise is equal to 1 and process noise is equal to 0.1. It is seen that the red colour line shows the Actual value of the arrival time, the blue colour line indicates the measured value of Arrival time and green colour line shows the estimated values of Arrival time at different intervals of time. Henece it is able to retrieve the state of node, from this case study, it is possible to retrieve the state and redirect the routing direction at the node level by using filter approach this leads optimal performace. For simuations, we used GNU OCTAVE TOOL and Ubutu Operating system 14.01.

## 6. CONCLUSION

Since the statistics are used in conjunction with the present state information, it is naturally expected that the Kalaman filter algorithm achieves better routing performance compared to earlier adaptive RWA algorithms especially M/M/1 under non-uniform traffic conditions, namely M/M/1. We suggest that further research in this direction is likely to find the over-head time taken to estimate arrival time and customer waitng time in each case and their comparisons with markov model may yield better result.

The effectiveness of the kalman filter approach for RWA problem compared to LLR and Max – sum routing approach will be communicated.

## AUTHORS

**Mohan Kumar S**. received his Bachelor of Engineering in Computer Science & Engineering and Master of .Technology, in Networking and Internet Engineering, from Visvesvaraya Technological University, Belgaum, and Karnataka, India respectively. He is currently working towards a Doctoral Degree from Visvesvaraya Technological University, Belgaum, and Karnataka, India. At present he is working as Assistant Professor, Department of Information Science and Engineering M.S.Ramaiah Institute of Technology Bangalore. Karnataka. India  (Affiliated to Visvesvaraya Technological University Belgaum),

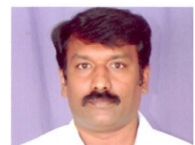

**Dr. Jagadeesha S N** received his Bachelor of Engineering., in Electronics and Communication Engineering, from University B. D. T College of Engineering., Davangere affiliated to Mysore University, Karnataka, India in 1979, M.E. from Indian Institute of Science (IISC), Bangalore, India specializing in Electrical Communication Engineering., in 1987 and Ph.D. in Electronics and Computer Engineering., from University of Roorkee, Roorkee, India in 1996. He is an IEEE member. His research interest includes Array Signal Processing, Wireless Sensor Networks and Mobile Communications. He has published and presented many papers on Adaptive Array Signal Processing and Direction-of-Arrival estimation. Currently he is professor in the Department of Computer Science and Engineering, Jawaharlal Nehru National College of Engineering. (Affiliated to Visvesvaraya Technological University), Shimoga, Karnataka, India

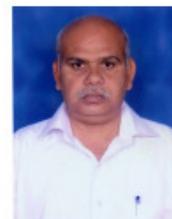